%% file: main.tex
\documentclass[manuscript]{acmart}
\makeatletter
\renewcommand\@formatdoi[1]{\ignorespaces}
\makeatother

\usepackage[ruled]{algorithm2e} 

\newcommand{\bomega}{\boldsymbol{\omega}}

\newcommand{\bdelta}{\boldsymbol{\delta}} 
 
\usepackage{tikz}
\usetikzlibrary{shapes,decorations,arrows,calc,arrows.meta,fit,positioning,decorations.pathmorphing}
\tikzset{
    -Latex,auto,node distance =1 cm and 1 cm,semithick,
    state/.style ={ellipse, draw, minimum width = 0.7 cm},
    point/.style = {circle, draw, inner sep=0.04cm,fill,node contents={}},
    bidirected/.style={Latex-Latex,dashed},
    el/.style = {inner sep=2pt, align=left, sloped}
}

\usepackage{multirow}
\usepackage{flushend}
\usepackage{amsmath,amssymb}
\usepackage{mathtools}
\usepackage[acronym]{glossaries}
\usepackage{units}
\usepackage{booktabs}
\usepackage[inline]{enumitem} 
\usepackage{subcaption}

\usepackage{arydshln}
\makeatletter
\def\adl@drawiv#1#2#3{%
        \hskip.5\tabcolsep
        \xleaders#3{#2.5\@tempdimb #1{1}#2.5\@tempdimb}%
                #2\z@ plus1fil minus1fil\relax
        \hskip.5\tabcolsep}
\newcommand{\cdashlinelr}[1]{%
  \noalign{\vskip\aboverulesep
           \global\let\@dashdrawstore\adl@draw
           \global\let\adl@draw\adl@drawiv}
  \cdashline{#1}
  \noalign{\global\let\adl@draw\@dashdrawstore
           \vskip\belowrulesep}}
\makeatother

\setcopyright{none}

\acmConference[REVEAL '20]{the ACM RecSys Workshop on Reinforcement Learning and Robust Estimators for Recommendation Systems}{}{}
\acmYear{2020}
\copyrightyear{2020}

\begin{document}
\title[]{From Clicks to Conversions: Recommendation for long-term reward}

\author{Philom\`ene Chagniot}
\affiliation{
  \institution{ENS Paris Saclay \& Criteo}
}
\email{philomene.chagniot@gmail.com}

\author{Flavian Vasile}
\affiliation{
  \institution{Criteo}
}
\email{f.vasile@criteo.com}

\author{David Rohde}
\affiliation{
  \institution{Criteo}
}
\email{d.rohde@criteo.com}

\renewcommand{\shortauthors}{}

\begin{abstract}
Recommender systems are often optimised for short-term reward: a recommendation is considered successful if a reward (e.g. a click) can be observed immediately after the recommendation. The advantage of this framework is that with some reasonable (although questionable) assumptions, it allows familiar supervised learning tools to be used for the recommendation task. However, it means that long-term business metrics, e.g. sales or retention are ignored. In this paper we introduce a framework for modeling long-term rewards in the RecoGym simulation environment. 
We use this newly introduced functionality to showcase problems introduced by the last-click attribution scheme in the case of conversion-optimized recommendations and propose a simple extension that leads to state-of-the-art results.
\end{abstract}

%
%

\maketitle

\input{1.intro}

\input{2.recogym_update}

\input{3.relaxed_last_click}

\bibliographystyle{ACM-Reference-Format}
\bibliography{bibliography}

\end{document}

%% file: 1.intro.tex
\section{Introduction}
A production recommender system produces logs of user timelines containing information about user behaviour, recommendations, responses to recommendation (e.g. clicks) and some notion of long-term reward (e.g. sales).  A modern approach to recommendation will look at this log in order to improve future recommendations. By examining how similar users respond to different recommendations it becomes possible to discover better recommendations and continue to improve the system.

This procedure of learning by experimentation in some respects mimics randomized control trials in medicine where populations are split into two and different treatments are delivered to similar groups.  Medical trials are however simpler, as an intervention or a placebo is administered to each group and then long-term impacts are observed with no further interventions delivered.
\paragraph{The challenges of credit attribution in the case of delayed reward and multiple actions}
In contrast with medical trials, where the treatment is frequently a binary variable, recommender systems will deliver multiple actions at variable times leading to combinatorially complex treatments. 
For simplicity, in our previous work on RecoGym \cite{rohde2018recogym}, we assumed that both the current recommendation and the reward are conditionally independent on past actions, therefore making the recommendation amenable to contextual bandits and supervised value modeling approaches. 

The problem of identifying the causal impact on the final reward of playing an action in a certain context from logged interactions was first outlined in \cite{sutton1985temporal} under the name of \emph{Temporal Credit Assignment}. In the computational advertising literature, a restrictive version of the same problem is known as the \emph{Conversion Attribution} problem.

Conversion-optimized recommendation is an active research area, but in most cases it skips the credit assignment problem, by borrowing ideas from the bidding approaches and using mostly last-click sales attribution, such as the work by \cite{Mann2019, wen2019conversion, wu2018turning}. One notable exception is the recent work by \cite{Xin2020}, which formulates the problem in the Reinforcement Learning (RL) framework. 

In our current work we remove this simplifying assumption, allowing recommendations to change the user state and therefore transforming the RecoGym environment into a full RL simulator. Using the newly introduced functionality we relax the overly-restrictive assumptions of the current last-click attribution scheme and show that under the new attribution, our recommendation agent outperforms all other state-of-the-art bandit baselines.
Overall, we believe that this paper makes two important contributions: firstly, we describe a simulation environment that enables study of long-term reward and we validate that it has sensible properties; 
secondly, we propose a novel reward attribution scheme that shows encouraging results in a variety of setups.

%% file: 2.recogym_update.tex
\section{A RecoGym extension for conversions modelling}
We consider an extension of the RecoGym environment \cite{rohde2018recogym}, where conversion (sale) events can happen on top of "organic" (views) and "bandit" (clicks) events. 

In standard RecoGym, the user state is constant and each product is represented by two different embeddings that represent the organic user behavior and the click-bandit behavior. Thus, in this setting, the recommendation problem can be framed as contextual bandits. We extend RecoGym by adding a further conversion embedding and more importantly allow for the hidden user state to be modified by recommendations.  

More specifically, a given product $a$ is now also defined by a conversion embedding ($\Lambda_{a}\in\mathbf{R}^{K}$). Each user is represented by conversion features ($\bdelta \in\mathbf{R}^{K}$) on top of organic ones ($\bomega \in\mathbf{R}^{K}$). At the beginning of the episode, the organic and conversion features of the user are fully aligned ($\bdelta=\bomega$). It will diverge along the episode based on the system's interventions.

We only modify the user's hidden state if the user clicks on a recommendations. After a click for product $a$, the \textit{user} conversion features $\bdelta$ are updated as $(1-\kappa)\bdelta+ \kappa \Lambda_{a}$, where $\kappa$ is a parameter between 0 and 1 and $\Lambda$ is the \textit{product} conversion features. Furthermore, we model sales as a $Bernoulli$ distribution whose mean is an increasing transformation of the dot product between $\bdelta$ and $\Lambda_{a}$.

Thus, ad displays and the resulting clicks will make the user state deviate from its initial organic value, and thus have a positive impact on the probability of sale.  


%% file: 3.relaxed_last_click.tex
\section{Shortcomings of last-click attribution for the case of conversion-optimized recommendation}
Last-click conversion attribution is a frequently-used attribution scheme used in the advertising industry as a heuristic for distributing value to various advertising channels that are being used across the same campaign. While a useful and easy way to bill for services, this attribution scheme is not a perfect reward shaping mechanism for conversion modeling, because it introduces a bias towards rewarding clicks that are simply \emph{correlated} with conversions, but that might not be \emph{causing} conversions. In our work, we introduce a causal graph that allows for conversions that are unmediated by ad clicks and show that under the assumption of strict post-click recommendation incrementality, a simple discounting mechanism of attributed sales can lead to better ranking of incremental actions. We will add all of the supporting experiments in our upcoming full paper.

%% file: main.bbl

\begin{thebibliography}{6}


\ifx \showCODEN    \undefined \def \showCODEN     #1{\unskip}     \fi
\ifx \showDOI      \undefined \def \showDOI       #1{#1}\fi
\ifx \showISBNx    \undefined \def \showISBNx     #1{\unskip}     \fi
\ifx \showISBNxiii \undefined \def \showISBNxiii  #1{\unskip}     \fi
\ifx \showISSN     \undefined \def \showISSN      #1{\unskip}     \fi
\ifx \showLCCN     \undefined \def \showLCCN      #1{\unskip}     \fi
\ifx \shownote     \undefined \def \shownote      #1{#1}          \fi
\ifx \showarticletitle \undefined \def \showarticletitle #1{#1}   \fi
\ifx \showURL      \undefined \def \showURL       {\relax}        \fi
\providecommand\bibfield[2]{#2}
\providecommand\bibinfo[2]{#2}
\providecommand\natexlab[1]{#1}
\providecommand\showeprint[2][]{arXiv:#2}

\bibitem[\protect\citeauthoryear{Mann, Gowal, Gyorgy, Hu, Jiang,
  Lakshminarayanan, and Srinivasan}{Mann et~al\mbox{.}}{2019}]%
        {Mann2019}
\bibfield{author}{\bibinfo{person}{Timothy~Arthur Mann}, \bibinfo{person}{Sven
  Gowal}, \bibinfo{person}{Andras Gyorgy}, \bibinfo{person}{Huiyi Hu},
  \bibinfo{person}{Ray Jiang}, \bibinfo{person}{Balaji Lakshminarayanan}, {and}
  \bibinfo{person}{Prav Srinivasan}.} \bibinfo{year}{2019}\natexlab{}.
\newblock \showarticletitle{Learning from Delayed Outcomes via Proxies with
  Applications to Recommender Systems}. In
  \bibinfo{booktitle}{\emph{Proceedings of the 36th International Conference on
  Machine Learning}} \emph{(\bibinfo{series}{Proceedings of Machine Learning
  Research})}, \bibfield{editor}{\bibinfo{person}{Kamalika Chaudhuri} {and}
  \bibinfo{person}{Ruslan Salakhutdinov}} (Eds.), Vol.~\bibinfo{volume}{97}.
  \bibinfo{publisher}{PMLR}, \bibinfo{address}{Long Beach, California, USA},
  \bibinfo{pages}{4324--4332}.
\newblock
\urldef\tempurl%
\url{http://proceedings.mlr.press/v97/mann19a.html}
\showURL{%
\tempurl}


\bibitem[\protect\citeauthoryear{Rohde, Bonner, Dunlop, Vasile, and
  Karatzoglou}{Rohde et~al\mbox{.}}{2018}]%
        {rohde2018recogym}
\bibfield{author}{\bibinfo{person}{David Rohde}, \bibinfo{person}{Stephen
  Bonner}, \bibinfo{person}{Travis Dunlop}, \bibinfo{person}{Flavian Vasile},
  {and} \bibinfo{person}{Alexandros Karatzoglou}.}
  \bibinfo{year}{2018}\natexlab{}.
\newblock \showarticletitle{Recogym: A reinforcement learning environment for
  the problem of product recommendation in online advertising}.
\newblock \bibinfo{journal}{\emph{arXiv preprint arXiv:1808.00720}}
  (\bibinfo{year}{2018}).
\newblock


\bibitem[\protect\citeauthoryear{Sutton}{Sutton}{1985}]%
        {sutton1985temporal}
\bibfield{author}{\bibinfo{person}{Richard~S Sutton}.}
  \bibinfo{year}{1985}\natexlab{}.
\newblock \showarticletitle{Temporal Credit Assignment in Reinforcement
  Learning.}
\newblock  (\bibinfo{year}{1985}).
\newblock


\bibitem[\protect\citeauthoryear{Wen, Zhang, Wang, Bao, Lin, and Yang}{Wen
  et~al\mbox{.}}{2019}]%
        {wen2019conversion}
\bibfield{author}{\bibinfo{person}{Hong Wen}, \bibinfo{person}{Jing Zhang},
  \bibinfo{person}{Yuan Wang}, \bibinfo{person}{Wentian Bao},
  \bibinfo{person}{Quan Lin}, {and} \bibinfo{person}{Keping Yang}.}
  \bibinfo{year}{2019}\natexlab{}.
\newblock \showarticletitle{Conversion Rate Prediction via Post-Click Behaviour
  Modeling}.
\newblock \bibinfo{journal}{\emph{arXiv preprint arXiv:1910.07099}}
  (\bibinfo{year}{2019}).
\newblock


\bibitem[\protect\citeauthoryear{Wu, Hu, Hong, and Liu}{Wu
  et~al\mbox{.}}{2018}]%
        {wu2018turning}
\bibfield{author}{\bibinfo{person}{Liang Wu}, \bibinfo{person}{Diane Hu},
  \bibinfo{person}{Liangjie Hong}, {and} \bibinfo{person}{Huan Liu}.}
  \bibinfo{year}{2018}\natexlab{}.
\newblock \showarticletitle{Turning clicks into purchases: Revenue optimization
  for product search in e-commerce}. In \bibinfo{booktitle}{\emph{The 41st
  International ACM SIGIR Conference on Research \& Development in Information
  Retrieval}}. \bibinfo{pages}{365--374}.
\newblock


\bibitem[\protect\citeauthoryear{Xin, Karatzoglou, Arapakis, and Jose}{Xin
  et~al\mbox{.}}{2020}]%
        {Xin2020}
\bibfield{author}{\bibinfo{person}{Xin Xin}, \bibinfo{person}{Alexandros
  Karatzoglou}, \bibinfo{person}{Ioannis Arapakis}, {and}
  \bibinfo{person}{Joemon~M. Jose}.} \bibinfo{year}{2020}\natexlab{}.
\newblock \bibinfo{title}{Self-Supervised Reinforcement Learning for
  Recommender Systems}.
\newblock
\newblock
\showeprint[arxiv]{cs.LG/2006.05779}


\end{thebibliography}
